\documentclass[journal,twoside,web]{ieeecolor}

\usepackage{ieeejbhi}
\usepackage{framed,multirow}
\usepackage{comment}

\usepackage{cite}
\usepackage{soul}
\usepackage{booktabs}
\usepackage{amsmath,amssymb,amsfonts}
\usepackage{algorithmic}
\usepackage[ruled,vlined]{algorithm2e}
\usepackage{graphicx}
\usepackage{textcomp}
\usepackage{placeins}
\usepackage{pifont} 
\usepackage{float}
\usepackage{array}
\usepackage{adjustbox}
\usepackage{bbm}
\usepackage{mathtools}
\usepackage{siunitx}
\usepackage{xcolor}
\usepackage{caption}

\SetKwInput{KwInput}{Input}                
\SetKwInput{KwOutput}{Output}              

\usepackage{subfig}
\usepackage{tikz}
\usetikzlibrary{matrix,calc}
\usepackage{amssymb}
\usepackage{latexsym}
\usepackage{gensymb}
\usepackage{url}
\usepackage{lipsum}
\usepackage{hyperref}
\hypersetup{
    colorlinks=true,
    linkcolor=blue,
    filecolor=magenta,      
    urlcolor=cyan,
    pdftitle={I2IMamba},
    }

\usepackage{xr}
\renewcommand{\arraystretch}{1.2}
\newcommand{\Tone}{T\textsubscript{1}}
\newcommand{\Ttwo}{T\textsubscript{2}}

\newcommand{\TtwoPD}{T\textsubscript{2}\,$\rightarrow$\,PD}

\newcommand{\PDTtwo}{PD\,$\rightarrow$\,T\textsubscript{2}}
\newcommand{\TtwoPDTone}{T\textsubscript{2},\,PD\,$\rightarrow$\,T\textsubscript{1}}
\newcommand{\TonePDTtwo}{T\textsubscript{1},\,PD\,$\rightarrow$\,T\textsubscript{2}}

\newcommand{\ToneTtwoPD}{T\textsubscript{1},\,T\textsubscript{2}\,$\rightarrow$\,PD}
\newcommand{\TtwoFlair}{T\textsubscript{2}\,$\rightarrow$\,FLAIR}
\newcommand{\FlairTtwo}{FLAIR\,$\rightarrow$\,T\textsubscript{2}}
\newcommand{\TtwoFlairTone}{T\textsubscript{2},\,FLAIR\,$\rightarrow$\,T\textsubscript{1}}
\newcommand{\ToneFlairTtwo}{T\textsubscript{1},\,FLAIR\,$\rightarrow$\,T\textsubscript{2}}
\newcommand{\ToneTtwoFlair}{T\textsubscript{1},\,T\textsubscript{2}\,$\rightarrow$\,FLAIR}

\makeatletter
\newcommand*{\addFileDependency}[1]{
  \typeout{(#1)}
  \@addtofilelist{#1}
  \IfFileExists{#1}{}{\typeout{No file #1.}}
}
\makeatother

\newcommand*{\myexternaldocument}[1]{%
    \externaldocument{#1}%
    \addFileDependency{#1.tex}%
    \addFileDependency{#1.aux}%
}

\myexternaldocument{I2IMamba_supp.tex}

\definecolor{newcolor}{rgb}{.8,.349,.1}
\definecolor{brightcerulean}{rgb}{0.11, 0.62, 0.74}

\usepackage[font=footnotesize,justification=justified,belowskip=2pt,aboveskip=2pt]{caption}
\begin{document}
\title{\vspace{-0.5cm}I2I-Mamba: Multi-modal medical image synthesis via selective state space modeling}
\author{Omer F. Atli, Bilal Kabas, Fuat Arslan, Arda C. Demirtas, Mahmut Yurt, Onat Dalmaz, \\and Tolga Çukur,~\IEEEmembership{Senior~Member}\vspace{-1.3cm}
  \thanks{Corresponding author: Tolga Çukur (e-mail: cukur@ee.bilkent.edu.tr). This work was supported in part by TUBA GEBIP 2015 and BAGEP 2017 fellowships awarded to T. \c{C}ukur, and in part by The Scientific and Technological Research Council of Türkiye (TÜBİTAK) 1515 Frontier R\&D Laboratories Support Program for Türk Telekom 6G R\&D Lab under project number 5249902.}%
  \thanks{O.F. Atli, B. Kabas, F. Arslan, A.C. Demirtas and T. Çukur are with the Department of Electrical and Electronics Engineering, and National Magnetic Resonance Research Center, Bilkent University, Ankara 06800, Turkey. M. Yurt and O. Dalmaz are with the Department of Electrical Engineering, Stanford University, CA 94305, United States.}%
}

\maketitle

\begin{abstract}
Multi-modal medical image synthesis involves nonlinear transformation of tissue signals between source and target modalities, where tissues exhibit contextual interactions across diverse spatial distances. As such,  the utility of a network architecture in synthesis depends on its ability to express the broad set of contextual features in medical images. Convolutional neural networks (CNNs) offer high local precision at the expense of poor sensitivity to long-range context. While transformers promise to alleviate this issue, they suffer from an unfavorable trade-off between sensitivity to long- versus short-range context due to the intrinsic complexity of attention filters. To effectively capture contextual features while avoiding the complexity-driven trade-offs, here we introduce a novel multi-modal synthesis method, I2I-Mamba, based on the state space modeling (SSM) framework. Focusing on high-level representations across a hybrid residual architecture, I2I-Mamba leverages novel dual-domain Mamba (ddMamba) blocks for complementary contextual modeling in image and Fourier domains, while maintaining spatial precision with convolutional layers. Diverting from conventional raster-scan trajectories, ddMamba leverages novel SSM operators based on a spiral-scan trajectory to learn context with enhanced angular isotropy and radial coverage, and a channel-mixing layer to aggregate context across the channel dimension. Comprehensive demonstrations on multi-contrast MRI and MRI-CT protocols indicate that I2I-Mamba outperforms state-of-the-art CNNs, transformers and SSMs.
\end{abstract}


\vspace{-0.3cm}
\bstctlcite{IEEEexample:BSTcontrol}

\section{Introduction}
Multi-modal medical images with distinct tissue contrasts provide complementary information about underlying anatomy, boosting reliability in downstream analyses \cite{pichler2008}. Multi-modal imaging is viable using different sequences on the same scanner or on entirely different scanners, albeit costs of running prolonged exams yield incomplete protocols under many scenarios \cite{thukral2015}. As a remedy, target-modality images missing from a protocol can be synthesized based on the subset of source-modality images available \cite{iglesias2013,huo2018}. Key clinical applications of image synthesis include imputing target modalities with diagnostically-relevant albeit redundant information that are omitted from imaging protocols to reduce scan time and improve efficiency; inferring invasive target modalities from non-invasive sources to avoid exposure to ionizing radiation or harmful contrast agents \cite{lee2019}; and recovering missing modalities to enhance protocol consistency across participants in retrospective imaging studies \cite{divbar2019}. Yet, such synthesis tasks involve a challenging nonlinear transformation of signal levels between source and target images depending on tissue characteristics \cite{lee2017}. Although detailed tissue parameters that govern signal levels are generally difficult to infer from medical images, multi-modal synthesis can be guided via a rudimentary spatial prior on tissue composition that can be implicitly inferred from the signal distribution in source images \cite{huang2017}. Note that both healthy and pathological tissues can exhibit broad spatial distribution in the form of contiguous or segregated clusters across an anatomy \cite{adam2014grainger}, introducing not only local signal correlations in compact neighborhoods but also non-local signal correlations over extended distances. Thus, successful solution of a multi-modal synthesis task inherently rests on the ability to exploit these short- to long-range contextual features.  

Deep learning has recently emerged as the mainstream framework for medical image synthesis given its prowess in nonlinear function approximation \cite{vemulapalli2015,wu2016,alexander2014,huynh2015}. In learning-based synthesis, a generative model attempts to map source onto target images through hierarchical nonlinear transformations on intermediate feature maps \cite{zhao2017,chartsias2018}. Naturally, the fidelity of this mapping depends on the model's expressiveness for the diverse set of contextual features in medical images. While open questions remain about optimal training strategies (e.g., adversarial vs. diffusion) \cite{chartsias2017,woltering2017 ,syndiff,MONAI}, architectural design is an independent and critical factor that directly determines the expressiveness of generative models. In this regard, earlier studies have used CNN-based models with convolution operators for local filtering of feature maps \cite{bowles2016,chartsias2018,joyce2017,yang2018,wei2019}. The popularity of CNNs has been fueled by linear model complexity with respect to image dimensionality, and high expressiveness for local context that can be critical in synthesis of detailed tissue structure \cite{beers2018,pgan,yu2018,yu2019ea,nie2018,armanious2019,lee2019,li2019,zhou2020,wang2020,yurt2021mustgan}. However, convolution operators have strictly localized spatial footprints, inducing poor sensitivity to long-range contextual features \cite{ptnet,ssdiffrecon}. In turn, CNNs can suffer from low synthesis accuracy in regions of heterogeneous tissue composition and uncommon pathology, where contextual relations are key in inferring the spatial distribution of tissue signals \cite{attention_unet,lan2020}. 

Later studies have instead adopted transformer-based models based on self-attention operators that are capable of non-local filtering \cite{ganbert,ptnet,resvit,gregtrans,trans_synth_shen}. Tokenizing an input image as a sequence of patches, transformers compute attention weights to measure inter-patch similarity and non-locally filter the input sequence. While the diffuse spatial footprints of self-attention operators help increase sensitivity to long-range context, they induce quadratic model complexity with respect to sequence length (i.e., number of image patches), prohibiting their use on small patches necessary to maintain high spatial precision \cite{gregtrans,trans_synth_shen}. Common strategies to facilitate their use include efficient attention operators that compress contextual representations at the expense of limiting contextual sensitivity \cite{ptnet,BolT}, and tokenization via large patches (e.g., a 16$\times$16 image patch taken as a single token) that compromises short-range contextual sensitivity \cite{resvit,gregtrans}. As such, transformer-based methods typically face an undesirable trade-off between spatial precision and global awareness. An emerging alternative to efficiently capture contextual representations is state-space models (SSM) \cite{zhu2024vision,liu2024vmamba}. SSMs cast image pixels onto an input sequence via raster-scan trajectories in rectilinear orientations, and efficiently model this sequence via state-space operators that offer a refined trade-off between long- and short-range contextual sensitivity. Recent reports have already adopted selective SSMs (Mamba) for unimodal analysis and reconstruction tasks in medical imaging \cite{ma2024UMamba,mambamir,mambarecon}. 

Given their success in other imaging tasks, SSMs hold significant promise for multi-modal medical image synthesis. However, several limitations hinder the effectiveness of existing SSMs in missing modality imputation. \textit{Traditional SSMs perform recurrent sequence modeling directly in the image domain} to aggregate contextual information, which often leads to an inevitable trade-off between capturing short- and long-range dependencies—\textit{global context is attenuated} due to the difficulty of modeling interactions between distant sequence elements \cite{ma2024UMamba}. Moreover, multi-modal images of the same anatomy typically exhibit \textit{signal correlations that vary with spatial frequency}: high-frequency edge structures are typically well-aligned across modalities, while low-frequency contrast information tends to be more modality-specific \cite{Bilgic2011MultiContrast}. Conventional image-domain SSMs can \textit{struggle to model this frequency-dependent behavior} effectively, limiting their ability to capture cross-modal interactions. Finally, SSMs typically impose \textit{a rectilinear raster-scan trajectory (e.g., sweep, zig-zag)} to sequentialize images, resulting in anisotropic receptive fields and \textit{reduced sensitivity to contextual interactions along non-axial directions} \cite{mambarecon}. Note that diagnostically relevant structures such as vasculature, cortical folds, and tumor lesions often follow oblique orientations that are poorly captured by rectilinear scans. These limitations collectively constrain the potential of SSMs in medical image synthesis tasks.

\begin{figure*}[!t]
\centerline{\includegraphics[width=0.9\textwidth]{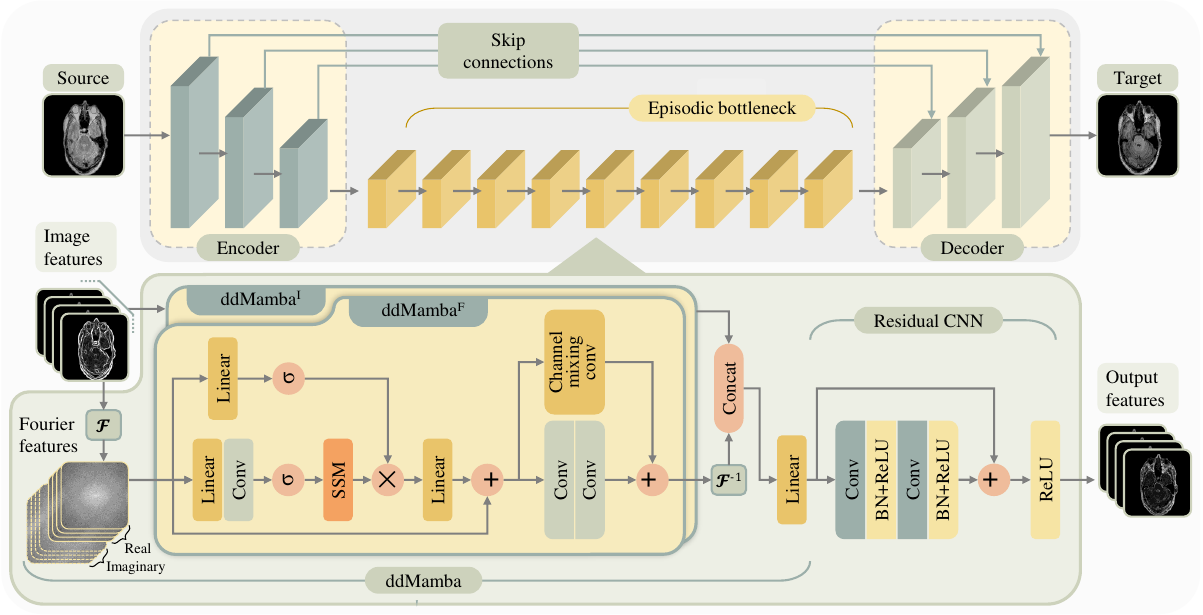}}
\caption{Network architecture for I2I-Mamba. The proposed model comprises encoder, bottleneck, and decoder modules to synthesize target from source images. The encoder extracts high-level representations of the source image via convolutional layers. The bottleneck extracts task-relevant contextual information across spatial, frequency and channel dimensions with the aid of dual-domain Mamba (ddMamba) blocks (ddMamba$^\text{I}$: image domain, ddMamba$^\text{F}$: Fourier domain) comprising channel-mixing layers, and maintains high spatial precision with the aid of residual CNN blocks. The decoder back-projects the contextualized representations onto the target image via convolutional layers.}
 \label{fig:main_fig}
\end{figure*}

In this study, we introduce I2I-Mamba, a novel learning-based model that, to our knowledge, represents the first SSM–based framework for multi-modal medical image synthesis (see \cite{atli_ismrm} for an earlier conference version employing conventional image-domain SSM blocks with raster-scan operators and focusing exclusively on multi-contrast MRI synthesis). To achieve high contextual sensitivity while addressing key limitations of existing SSM approaches, I2I-Mamba incorporates dual-domain Mamba (ddMamba) blocks, residually fused with CNN blocks across a bottleneck that preserves high-resolution semantic representations (Fig. \ref{fig:main_fig}). To overcome the limitations of conventional image-domain SSMs in capturing global context and frequency-dependent interactions, ddMamba blocks feature independent SSM branches operating in the image and Fourier domains. This dual-domain design enables the state-space operators to leverage both spatial and spectral contextual cues. To mitigate the anisotropic spatial footprint introduced by traditional raster-scan trajectories, ddMamba blocks adopt a novel spiral scan trajectory, enhancing both radial coverage and angular isotropy in receptive fields (Fig. \ref{fig:main_fig2}). Additionally, ddMamba blocks incorporate channel-mixing layers to enable context aggregation not only across spatial dimensions but also across channel dimensions of the feature maps. 

The architectural design of I2I-Mamba is devised to capture rich cross-modal contextual interactions in medical images. We demonstrate the effectiveness of this design through comprehensive experiments on missing modality imputation in multi-contrast MRI and MRI-to-CT translation tasks. To systematically evaluate the impact of architecture, primary comparisons are performed in an adversarial training setup, providing a computationally efficient framework for controlled evaluation, while additional experiments benchmark I2I-Mamba against recent diffusion baselines. Our results show that I2I-Mamba consistently outperforms state-of-the-art CNN, transformer, and SSM-based models. The code is publicly available at: \href{https://github.com/icon-lab/I2I-Mamba}{https://github.com/icon-lab/I2I-Mamba}.

\vspace{4mm}
\textbf{Contributions}
\begin{itemize}
    \item We develop, to our knowledge, the first application of SSMs for translating between multi-modal data to synthesize medical images from missing modalities.
    \item We propose a novel SSM model that operates on image and Fourier representations to enhance medical image synthesis by capturing complementary spatial and spectral features.
    \item We design a novel ddMamba block that modulates SSM outputs with channel-mixing layers, enabling the model to jointly learn contextual dependencies across spatial-spectral and channel dimensions.
    \item We introduce a spiral-scan trajectory for image- and Fourier-domain SSM layers that enhances radial coverage and angular isotropy in receptive fields, improving the modeling of context in non-axial directions.
\end{itemize}

\section{Related Work}
\subsection{Learning-based medical image synthesis}
Learning-based image synthesis methods rely on the choice of model architecture to infer mappings across modalities. CNNs have been a mainstream architecture, using convolution operators to extract local image context. However, they struggle to generalize to atypical anatomy and model long-range dependencies \cite{attention_unet}. Attention mechanisms have been considered in CNN backbones to emphasize semantically relevant regions during synthesis \cite{lan2020,zhao2020,yuan2020,SelfRDB}. Yet, because these mechanisms multiplicatively gate features derived via convolution operators, they provide only a modest change in sensitivity to global context \cite{resvit}. 

Transformers have been increasingly adopted to explicitly model long-range interactions between spatially distant regions \cite{ptnet}. While powerful, pure transformer-based models incur high computational costs, which has led to the development of efficient variants using approximations such as windowed or low-rank attention \cite{BolT}. These approximations, however, imit the model's ability to capture the full-scope of contextual interactions \cite{gregtrans}. Alternatively, hybrid CNN-transformer architectures apply transformer modules only in low-resolution stages to reduce computation \cite{trans_unet,resvit}, compromising local precision. Consequently, transformer-based methods often trade off local precision for broader context due to the inherent computational limitations of attention mechanisms.

\subsection{SSM models in medical imaging}
SSMs have recently emerged as promising alternatives for achieving a refined balance between short- and long-range sensitivity, without incurring high model complexity. In medical imaging, SSM models have recently been proposed for unimodal tasks such as image segmentation \cite{ma2024UMamba}, image reconstruction from undersampled measurements \cite{mambamir,mambarecon}, and image generation \cite{vmddpm}. However, these studies have not addressed the distinct challenges posed by multi-modal synthesis. Here we introduce I2I-Mamba, which, to the best of our knowledge, is the first SSM-based framework designed for mapping between distinct imaging modalities, including both multi-contrast MRI and MRI-to-CT translation. While few independent studies \cite{mamformer} have appeared on this topic following our preprint \cite{atlipreprint}, I2I-Mamba differs from these efforts through several architectural novelties, including (i) a residual SSM–CNN design —excluding transformer modules— that departs from the common UNet-style backbone to better preserve low-level spatial detail, (ii) dual-domain SSM blocks that operate jointly in image and frequency domains rather than conventional image-domain SSMs, and (iii) a spiral-scan tokenization strategy that replaces standard raster scans.

Common UNet-style SSMs pervasively adopted in the imaging literature typically reduce spatial resolution to coarse feature maps (e.g., 16×16) \cite{liu2024vmamba}, thereby limiting spatial precision in contextual modeling \cite{pgan}. Furthermore, conventional SSM models often operate solely in the image domain, making them less effective at capturing frequency-dependent patterns and long-range global structures. These models also use raster-scan trajectories to map images into 1D sequences, which introduces anisotropic biases in horizontal and vertical directions \cite{liu2024vmamba}. I2I-Mamba departs from existing SSM architectures in imaging in several ways. First, it employs a deep latent bottleneck at higher resolution (64×64), preserving local spatial detail while enabling efficient contextualization. Second, it uses dual-domain SSM processing, jointly operating in both image and Fourier domains. This dual representation enables state-space operators to simultaneously learn spatially localized features and globally coherent spectral patterns, offering improved sensitivity to multi-scale and frequency-aware features. Third, we propose a spiral-scan trajectory for SSM tokenization, yielding near-isotropic receptive fields and mitigating directional bias in contextual modeling. These technical advances position I2I-Mamba as a performant solution for modality translation in medical imaging.


\begin{figure*}[t]
\centerline{\includegraphics[width=0.8\textwidth]{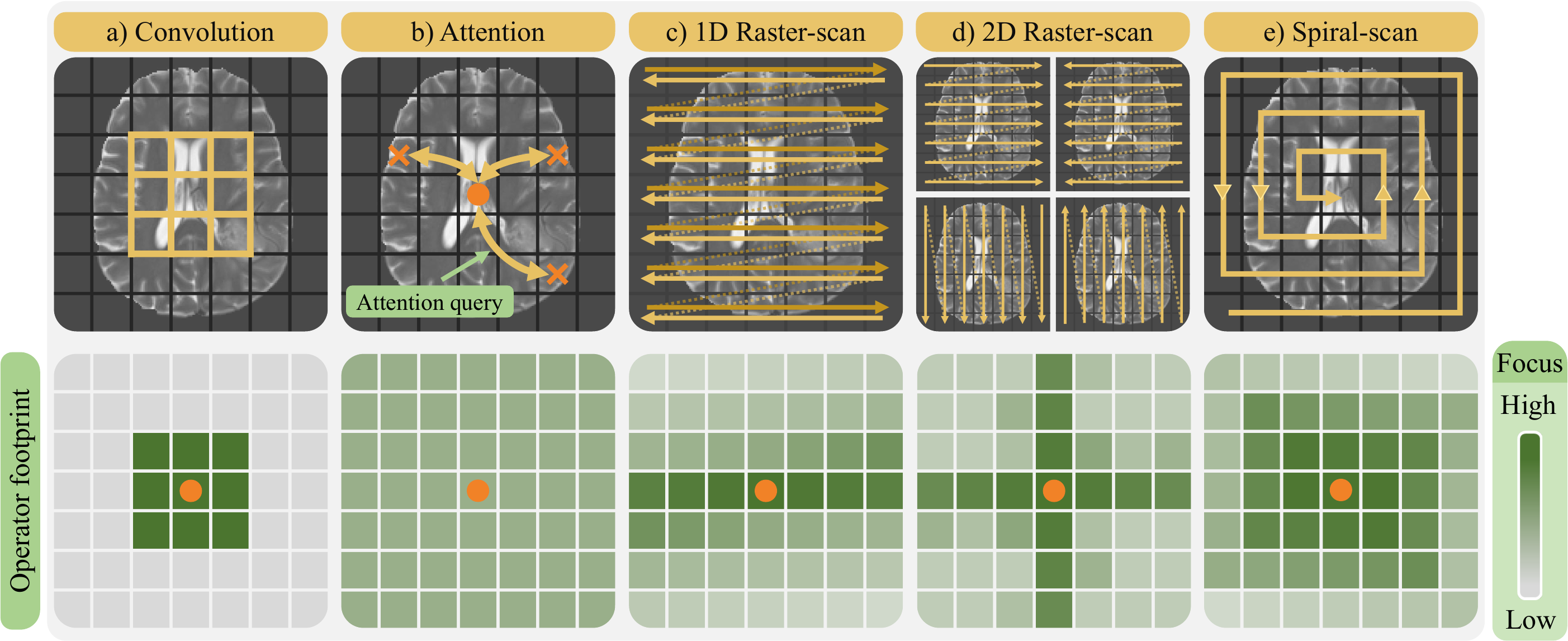}}
\caption{Footprints illustrating the spatial distribution of focus that each learning operator deploys (see colorbar), while seeking contextual interactions of a central pixel (orange dots). \textbf{(a)} Convolution operators in CNNs have localized footprints with heavy focus over a restricted neighborhood, compromising sensitivity to long-range contextual interactions. \textbf{(b)} Attention operators in transformers have non-local footprints that diffusely distribute focus over the image, compromising local precision. \textbf{(c), (d)} Conventional state-space operators in SSMs are based on multiple raster-scan trajectories with anisotropic footprints biased towards rectangular image axes, limiting sensitivity to interactions in non-axial orientations. \textbf{(e)} I2I-Mamba's state-space operator leverages a novel spiral-scan trajectory that attains a near-isotropic footprint with more uniform focus across orientations, maintaining an improve balance between long- versus short-range contextual interactions.}
 \label{fig:main_fig2}
\end{figure*} 

\subsection{Multi-domain processing in medical image synthesis}
Multi-domain processing has remained relatively underexplored in medical image synthesis. Existing methods largely operate in the image domain, with limited attention to complementary transform-based representations. A recent study has investigated the multi-resolution nature of wavelets to capture hierarchical image features \cite{Friedrich_2024}. However, the wavelet-domain method is based on a convolutional backbone, hence does not benefit from the contextual modeling capabilities of state-space operators in transform domains. To our knowledge, I2I-Mamba is the first to employ SSM in a dual-domain formulation based on image- and Fourier-domain representations. This dual-domain design enables context fusion across spatial and spectral representations, offering a principled way to integrate localized and global structural contents.

\section{Methods}

\subsection{I2I-Mamba}
I2I-Mamba is a novel medical image synthesis model that combines the contextual sensitivity of SSMs with the local precision of CNNs in a hybrid adversarial architecture. The generator of I2I-Mamba comprises encoder, high-resolution bottleneck, and decoder stages to synthesize target-modality images from source-modality images (Fig. \ref{fig:main_fig}). 

\textbf{Encoder.} Receiving as input available source-modality images, the encoder extracts high-resolution semantic representations after concatenating input modalities into a tensor:
\begin{align}
    X = [x_1; x_2; \dots; x_I],
\end{align}
where $I \in \mathbb{Z}^+$ is the number of source modalities, $x_i \in \mathbb{R}^{H,W}$ is the source image for the $i$th modality, and $X \in \mathbb{R}^{H,W,I}$ is the input tensor ($H$, $W$: image height and width). The input tensor is projected through multiple CNN blocks to derive latent representations at relatively high spatial resolution:
\begin{align}
    f_1 = \mathsf{CNN}(X),
\end{align}
where $f_1 \in \mathbb{R}^{H'=H/\alpha,W'=W/\alpha,C}$ ($\alpha$: moderate downsampling rate, $C$: the number of feature channels). 

\textbf{Dual-domain Mamba bottleneck.} Next, a deep bottleneck with $J$ stages extracts task-relevant contextual representations. At each stage, \underline{\textit{novel ddMamba blocks}} are employed that contain SSM layers independently operating in image and Fourier domains to capture spatial and spectral context (Fig.~\ref{fig:main_fig}), followed by channel-mixing layers to aggregate contextual interactions among the channel dimension. The contextualized feature maps from the ddMamba blocks are further processed via a residual CNN (rCNN) block to maintain high spatial precision. Receiving the feature map $f_j\in\mathbb{R}^{H',W',C}$, the $j$th bottleneck stage computes $f_{j+1}\in\mathbb{R}^{H',W',C}$ as:
\begin{align}
   f_{j+1} &= \mathsf{rCNN}(\mathsf{ddMamba}(f_j)).
\end{align}
In each ddMamba block, a first branch $\mathsf{ddMamba}^{I}$ derives feature map $e^I_{cm}\in\mathbb{R}^{H',W',C}$ operating in the image domain:
\begin{align}
    e^I_{cm} &= \mathsf{ddMamba^I}(f_j).
\end{align}
A second branch $\mathsf{ddMamba}^{F}$ derives feature map $e^F_{cm}\in\mathbb{R}^{H',W',C}$ operating in the Fourier domain:
\begin{align}
    e^F_{cm} &= \mathcal{F}^{-1}\{\mathsf{ddMamba^F}(\mathcal{F}\{f_j\})\},
\end{align}
where $\mathcal{F},\mathcal{F}^{-1}$ denote forward and inverse Fourier transformation, respectively. To process complex-valued feature maps in $\mathsf{ddMamba}^{F}$, real and imaginary components of $\mathcal{F}\{f_j\}$ are stacked across the channel dimension, projected through an SSM layer with shared weights, and recombined into complex-valued feature maps prior to inverse Fourier transformation to preserve the overall structure. The contextualized feature maps are concatenated and two-fold compressed in the channel dimension via a linear projection:
\begin{align}
\Tilde{e}_{cm} = \mathsf{Lin}([e^I_{cm}; \, e^F_{cm}]),
\end{align}
which is forwarded to the rCNN block \cite{resnet}. 

Within $\mathsf{ddMamba}$ branches, a gating variable $g\in\mathbb{R}^{H',W',C'}$ ($C'=\beta C$ for $\mathsf{ddMamba}^{I}$, $C'=2\beta C$ for $\mathsf{ddMamba}^{F}$) is first computed as:
\begin{align}
    g = \sigma ( \mathsf{Lin}_{\beta}(f_{j})),
\end{align}
where $\sigma$ is an activation function, $\mathsf{Lin}$ has expansion factor $\beta$. The input feature map $f_j$ (or $\mathcal{F}(f_j)$) is then embedded via depth-wise convolution, and passed through an SSM layer:
\begin{align}
    e_{j} =& \sigma(\mathsf{DWConv}(\mathsf{Lin}_{\beta}(f_{j}))),\\
    M =& \mathsf{SSM}(e_j).    \label{eq:SSMblock}
\end{align}

\textbf{Spiral-scan SSM operators.} In conventional SSM layers, the input feature map is expanded along spatial dimensions onto a 1D sequence via multiple raster-scan trajectories (e.g., based on rectilinear sweep or zig-zag patterns) \cite{liu2024vmamba}. This causes state-space operators to have anisotropic footprints biased towards rectangular image axes, which can compromise learning of contextual features in remaining orientations (Fig. \ref{fig:main_fig2}c-d). To alleviate orientation bias in state-space operators, here we propose a \underline{\textit{novel spiral-scan trajectory}} that maps $e_j\in\mathbb{R}^{H',W',C'}$ onto a sequence $z_{in}$ as (Fig. \ref{fig:main_fig2}e):
\begin{align}
\label{eq:spiral}
     &z_{in}[n,c] = e_j(h[n],w[n],c), \quad \mbox{such that:} \\
    &h[n] = (1-\Gamma)\lfloor \frac{H'+2}{2}\rfloor + \sum_{k=1}^{(H'W'-n+1)}\Gamma \mathrm{cos}(\frac{\pi}{2}\lfloor\sqrt{4k-7}\rfloor), \nonumber \\ 
    &w[n] =  (1-\Gamma)\lfloor \frac{W'+2}{2} \rfloor - \sum_{k=1}^{(H'W'-n+1)}\Gamma\mathrm{sin}(\frac{\pi}{2}\lfloor\sqrt{4k-7}\rfloor). \nonumber
\end{align}    
where $n\in\mathbb{Z}^{[1\mbox{ }H'W']}$ is the sequence index, $\Gamma=\mathrm{sgn}(k$-$1)$ with $\mathrm{sgn}$ denoting signum function, $\lfloor \cdot \rfloor$ is the floor operation, and ($h\in\mathbb{Z}^{[1\mbox{ }H']}$, $w\in\mathbb{Z}^{[1\mbox{ }W']}$) denote the ordering of rectangular pixel coordinates.

The sequence is then processed via the discretized state-space operator separately across channels:
\begin{align}
    \ell[n,c] &= \mathbf{\bar{A}} \ell[n-1,c] + \mathbf{\bar{B}}[n] z_{in}[n,c],\\
    z_{out}[n,c] &= \mathbf{\bar{C}}[n] \ell[n,c],
\end{align}
where $\ell$ is the hidden state, $\mathbf{\bar{A}} \in \mathbb{R}^{N,N}$ is a learnable and $\mathbf{\bar{B}}[n] \in \mathbb{R}^{N,1}$, $\mathbf{\bar{C}}[n] \in \mathbb{R}^{1,N}$ are learnable, input-dependent parameters, $N$ is the state dimensionality. By inverting the coordinate mapping in Eq. \ref{eq:spiral}, the output sequence $z_{out}$ is remapped onto feature map $M\in\mathbb{R}^{H',W',C'}$. After gating $M$ via a Hadamard product, it is linearly projected and residually combined with $f_j$:
\begin{align}
    e_{\mathrm{SSM}} = f_j+\mathsf{Lin}_{1/\beta}(g \odot M).
\end{align}
Note that $e_{\mathrm{SSM}}\in\mathbb{R}^{H',W',C'}$ primarily captures spatial or spectral relationships in feature maps, while treating channels independently. Thus, we propose to include a channel-mixing layer in ddMamba blocks to aggregate context across channels:
\begin{align}
    e_{cm} = \mathsf{cmConv}(e_{\mathrm{SSM}}) + \mathsf{Conv}(e_{\mathrm{SSM}}),
\end{align}
where channel mixing is achieved via a 1$\times$1 convolution operator \cite{mlp_mixer}. Next, $\Tilde{e}_{cm}\in\mathbb{R}^{H',W',C}$ fused across the ddMamba branches is projected through an rCNN block to compute the output feature map $f_{j+1}$.

\begin{table*}[t]
\centering
\caption{Performance for multi-contrast MRI synthesis tasks in IXI. PSNR (dB) and SSIM are listed as mean$\pm$std across the test set. Convolutional (Conv), diffusion (Diff), transformer (Trans), and state-space (SSM) baselines were considered. Boldface indicates the top-performing model for each task.}
\renewcommand{\arraystretch}{1.2}
\resizebox{\textwidth}{!}{%
\begin{tabular}{l@{\hspace{8pt}}cccccccccccc}
\hline
\multirow{2}{*}{\textbf{}} & \multirow{2}{*}{\textbf{}} & \multicolumn{2}{c}{\ToneTtwoPD} & \multicolumn{2}{c}{\TonePDTtwo} & \multicolumn{2}{c}{\TtwoPDTone} & \multicolumn{2}{c}{\TtwoPD} & \multicolumn{2}{c}{\PDTtwo} \\ 
\cmidrule(lr){3-4} \cmidrule(lr){5-6} \cmidrule(lr){7-8} \cmidrule(lr){9-10} \cmidrule(lr){11-12}
 &  & PSNR & SSIM & PSNR & SSIM & PSNR & SSIM & PSNR & SSIM & PSNR & SSIM \\ \hline
\multirow{3}{*}{\rotatebox[origin=c]{90}{\textit{Conv}}}
& medSynth\cite{nie2018}
& 30.69$\pm$2.43 & 0.940$\pm$0.019
& 32.66$\pm$2.40 & 0.963$\pm$0.017
& 27.10$\pm$2.18 & 0.915$\pm$0.037
& 29.10$\pm$2.44 & 0.909$\pm$0.013
& 30.41$\pm$2.08 & 0.956$\pm$0.016 \\ \cline{2-12}
& pGAN\cite{pgan}
& 31.40$\pm$2.40 & 0.948$\pm$0.013
& 31.96$\pm$1.95 & 0.955$\pm$0.015
& 27.14$\pm$2.45 & 0.928$\pm$0.031
& 30.86$\pm$2.39 & 0.957$\pm$0.015
& 30.11$\pm$2.34 & 0.949$\pm$0.019 \\ \cline{2-12}
& WDM\cite{Friedrich_2024}
& 32.50$\pm$2.48 & 0.966$\pm$0.011
& 33.30$\pm$1.75 & 0.964$\pm$0.012
& 28.01$\pm$2.48 & 0.934$\pm$0.033
& 31.27$\pm$2.48 & 0.960$\pm$0.014
& 32.34$\pm$1.55 & 0.961$\pm$0.012 \\ \hline
\multirow{2}{*}{\rotatebox[origin=c]{90}{\textit{Diff}}}
& DDPM\cite{NEURIPS2020_4c5bcfec}
& 31.77$\pm$5.31 & 0.893$\pm$0.076
& 32.50$\pm$2.17 & 0.948$\pm$0.049
& 26.41$\pm$4.37 & 0.921$\pm$0.077
& 29.68$\pm$6.39 & 0.720$\pm$0.255
& 31.87$\pm$2.97 & 0.964$\pm$0.024 \\ \cline{2-12}
& SynDiff\cite{syndiff}
& 30.86$\pm$5.21 & 0.934$\pm$0.093
& 29.35$\pm$8.45 & 0.890$\pm$0.087
& 26.23$\pm$2.66 & 0.918$\pm$0.038
& 29.85$\pm$4.86 & 0.930$\pm$0.092
& 28.42$\pm$6.09 & 0.925$\pm$0.111 \\ \hline
\multirow{4}{*}{\rotatebox[origin=c]{90}{\textit{Trans}}}
& PTNet\cite{ptnet}
& 30.33$\pm$2.59 & 0.935$\pm$0.018
& 32.62$\pm$2.09 & 0.954$\pm$0.013
& 27.73$\pm$2.79 & 0.931$\pm$0.029
& 29.75$\pm$2.66 & 0.932$\pm$0.012
& 32.46$\pm$2.19 & 0.956$\pm$0.022 \\ \cline{2-12}
& ResViT\cite{resvit}
& 32.98$\pm$2.35 & 0.968$\pm$0.027
& 33.47$\pm$2.54 & 0.964$\pm$0.014
& 28.45$\pm$1.46 & 0.936$\pm$0.011
& 32.08$\pm$2.49 & 0.952$\pm$0.010
& 32.84$\pm$1.56 & 0.964$\pm$0.011 \\ \cline{2-12}
& TransUNet\cite{trans_unet}
& 30.84$\pm$2.27 & 0.953$\pm$0.031
& 32.36$\pm$1.97 & 0.960$\pm$0.015
& 27.34$\pm$1.95 & 0.928$\pm$0.033
& 29.01$\pm$2.37 & 0.927$\pm$0.019
& 31.73$\pm$1.87 & 0.958$\pm$0.015 \\ \cline{2-12}
& Swin-Unet\cite{swinunet}
& 31.37$\pm$2.63 & 0.966$\pm$0.011
& 31.35$\pm$4.25 & 0.958$\pm$0.039
& 26.11$\pm$3.28 & 0.933$\pm$0.036
& 29.32$\pm$4.07 & 0.941$\pm$0.049
& 30.87$\pm$3.89 & 0.960$\pm$0.035 \\ \hline
\multirow{2}{*}{\rotatebox[origin=c]{90}{\textit{SSM}}}
& U-Mamba\cite{ma2024UMamba}
& 31.96$\pm$2.43 & 0.960$\pm$0.012
& 28.53$\pm$2.27 & 0.937$\pm$0.030
& 27.73$\pm$2.52 & 0.929$\pm$0.040
& 32.26$\pm$2.30 & 0.963$\pm$0.014
& 26.48$\pm$5.87 & 0.900$\pm$0.098 \\ \cline{2-12}
& Mamba-Unet\cite{mambaunet}
& 29.56$\pm$4.48 & 0.967$\pm$0.016
& 32.39$\pm$3.90 & 0.966$\pm$0.035
& 26.93$\pm$3.40 & 0.940$\pm$0.037
& 29.54$\pm$3.88 & 0.948$\pm$0.046
& 30.33$\pm$3.97 & 0.955$\pm$0.035 \\ \hline

& I2I-Mamba
& \textbf{33.46$\pm$2.52} & \textbf{0.969$\pm$0.011}
& \textbf{34.59$\pm$2.03} & \textbf{0.970$\pm$0.011}
& \textbf{29.15$\pm$2.48} & \textbf{0.947$\pm$0.027}
& \textbf{32.59$\pm$2.48} & \textbf{0.967$\pm$0.012}
& \textbf{33.89$\pm$1.65} & \textbf{0.968$\pm$0.011} \\ \hline
\end{tabular}
}
\label{tab:ixi}
\end{table*}

\textbf{Decoder.} The decoder receives contextualized feature maps $f_J$ from the bottleneck and projects them onto synthetic target image $y \in \mathbb{R}^{H,W}$ via transposed convolution blocks:
\begin{align}
   y = \mathsf{CNN_{transposed}}(f_J).
\end{align}
I2I-Mamba is implemented as an adversarial model with a patch-based discriminator $D$ \cite{nie2018}, distinguishing actual ($y_{act}$) and synthetic ($y_{syn}$) target images. A combined pixel-wise and adversarial objective is used to train the generator $G$ \cite{resvit}: 
\begin{align}
L_{G} =& \lambda_{pix}  \mathbb{E}[||y_{syn}-y_{act}||_1]  \quad - \nonumber \\ 
   & \lambda_{adv} \big\{ \mathbb{E}[D(y_{act}|X)^2] + \mathbb{E}[(D(y_{syn}|X)-1)^2] \big\},
\label{eq:genloss}
\end{align}
where $\mathbb{E}$ denotes expectation, $y_{syn}=G(X)$, $\lambda_{pix}$ and $\lambda_{adv}$ are loss-term weightings. Meanwhile, an adversarial term is used to train the discriminator $D$ \cite{nie2018}:
\begin{align}
L_{D} =& \mathbb{E}[D(y_{act}|X)^2] + \mathbb{E}[(D(y_{syn}|X)-1)^2].
\label{eq:disloss}
\end{align}

\begin{figure*}[t]
\centering
\includegraphics[width=0.725\textwidth]{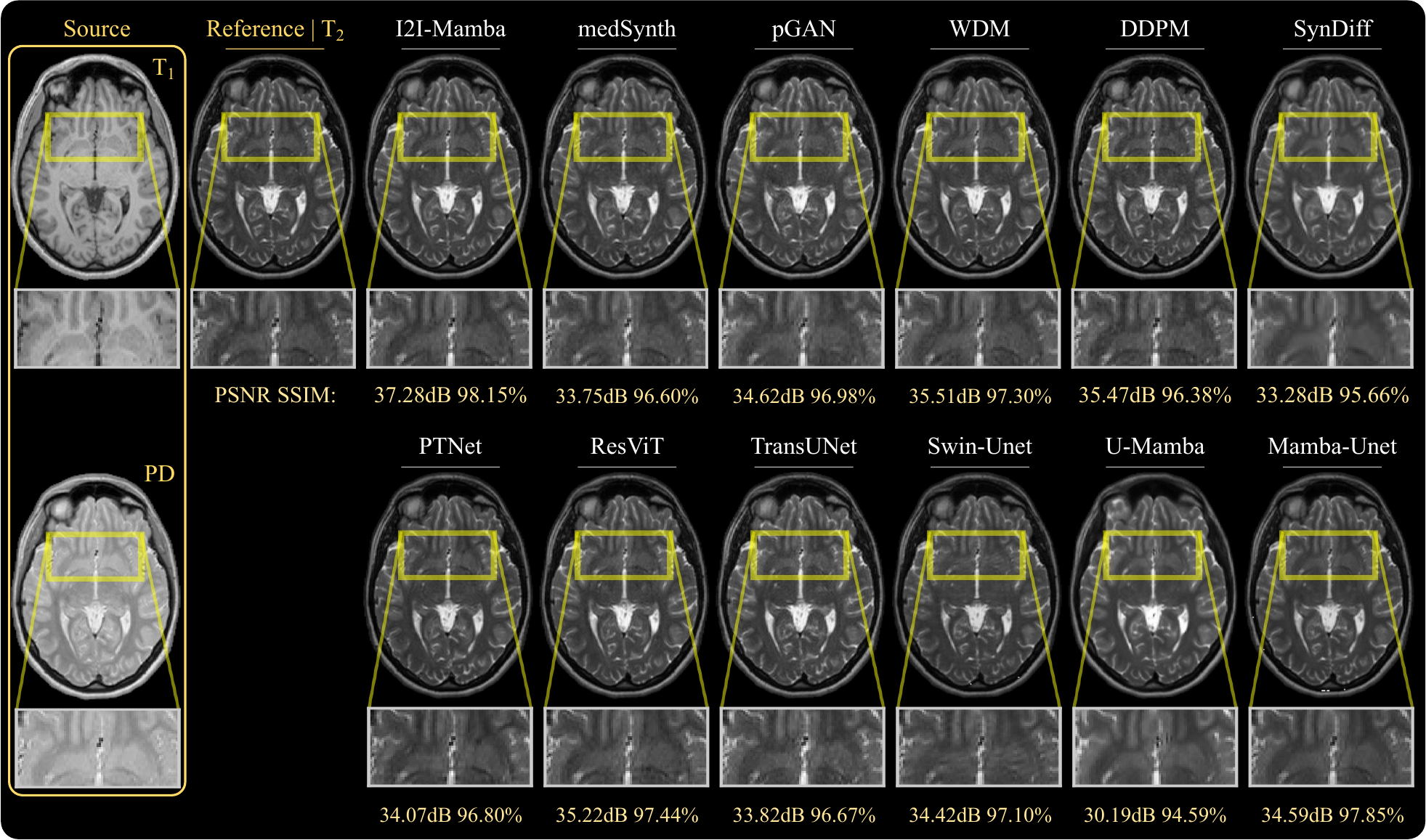}
\captionsetup{justification=justified,singlelinecheck=false}
\caption{Representative results for \TonePDTtwo~in IXI. Synthetic target images from competing methods are displayed along with source images and reference target images. Zoom-in windows and performance metrics are also included to highlight differences among methods.}
\label{fig:IXI}
\end{figure*}

\subsection{Datasets}
To mitigate potential biases due to demographic or protocol-related imbalances, we conducted evaluations on three distinct publicly available datasets with varying imaging protocols and subject populations. Experiments were conducted on two multi-contrast brain MRI datasets (IXI: \href{https://brain-development.org/ixi-dataset/}{https://brain-development.org/ixi-dataset/}, BraTS \cite{brats2021}) and a multi-modal pelvic MRI-CT dataset \cite{mr_ct_dataset}. For each dataset, samples were selected sequentially as released in the original repositories to avoid manual filtering that could introduce bias. Experiments confirmed that model performance converged beyond the selected training set sizes.

\textbf{IXI.} \Tone-, \Ttwo-, PD-weighted MR images were analyzed with ($7500$, $3000$, $5400$) cross-sections reserved for (training, validation, test) sets, based on a subject-level split. Prior to modeling, \Ttwo- and PD-weighted images were spatially registered onto \Tone-weighted images in each subject via an affine transformation in FSL. 

\textbf{BraTS.} \Tone-, \Ttwo-, FLAIR-weighted MR images from were analyzed with ($7500$, $3000$, $7500$) cross-sections reserved for (training, validation, test) sets, based on a subject-level split. As publicly shared, this dataset provides images that are co-registered onto \Tone-weighted MRI scans.

\textbf{MRI-CT.}
\Tone-, \Ttwo-weighted MRI, and CT images from were analyzed with a ($2430$, $540$, $1080$) cross-section reserved for (training, validation, test) sets, based on a subject-level split. As publicly shared, this dataset provides multi-modal images that are co-registered onto \Ttwo-weighted MRI scans.

\subsection{Architectural Design}
The encoder module in I2I-Mamba had 3 stages, each containing a CNN block with a convolutional layer, batch normalization (BN), and ReLU activation. $H$=256, $W$=256, $\alpha$=4, $C$=256 were used. The bottleneck had 9 stages, each containing a residual CNN block with 2 cascades of a convolutional layer, BN, and ReLU activation. Dual-domain ddMamba blocks were inserted in bottleneck stages $j$=\{1,5,9\}, and used $\beta$=2, $N$=16, and SiLU activations. Channel-mixing blocks used a parallel combination of a channel-mixing convolutional layer and 2 regular convolutional layers. The decoder module had 3 stages, each containing a CNN block with a convolutional layer, BN, and ReLU activation, except for the final stage that used a Tanh activation. Long-range skip connections were employed between corresponding encoder-decoder stages to better preserve low-level spatial information. 

\subsection{Competing Methods}
State-of-the-art baselines for medical image synthesis were considered including convolutional models (medSynth \cite{nie2018}, pGAN \cite{pgan}, WDM \cite{Friedrich_2024}), transformer models (PTNet \cite{ptnet}, ResViT \cite{resvit}, TransUNet \cite{trans_unet}, Swin-Unet \cite{swinunet}), and recent SSM models (U-Mamba \cite{ma2024UMamba}, Mamba-Unet \cite{mambaunet}). For systematic evaluations that clearly isolate the impact of architecture, these competing methods were all implemented as adversarial models that processed individual cross-sections, used a PatchGAN discriminator, and trained using the losses expressed in Eqs. \ref{eq:genloss}-\ref{eq:disloss}. To benchmark I2I-Mamba against recent generative frameworks, diffusion baselines were also considered (DDPM \cite{NEURIPS2020_4c5bcfec}, SynDiff \cite{syndiff}). Diffusion baselines were trained via procedures described in their original papers.

\subsection{Modeling Procedures}
Modeling was performed using the PyTorch framework on an Nvidia RTX 4090 GPU. All models were trained from scratch via the Adam optimizer with parameters $\beta_1=0.5,\,\beta_2=0.999$. Model hyperparameters were selected via stratified cross-validation. Data splitting was performed at the subject level, ensuring that no subject appeared in more than one fold. To avoid potential biases, a common set of hyperparameters observed to yield near-optimal validation performance across models was selected. Accordingly, $2\times10^{-4}$ learning rate, 60 epochs, $\lambda_{adv}=1,\lambda_{pix}=100$ were prescribed. To evaluate performance, peak signal-to-noise ratio (PSNR) and structural similarity index (SSIM) metrics were measured between ground-truth and synthetic images. Significance of performance differences was assessed with non-parametric signed-rank tests.


\begin{table*}[t]
\centering
\caption{Performance for multi-contrast MRI synthesis tasks in BraTS. PSNR (dB) and SSIM are listed as mean$\pm$std across the test set. Boldface indicates the top-performing model for each task.}
\renewcommand{\arraystretch}{1.2}
\resizebox{\textwidth}{!}{%
\begin{tabular}{l@{\hspace{8pt}}ccccccccccc}
\hline
\multirow{2}{*}{\textbf{}} & \multirow{2}{*}{} & \multicolumn{2}{c}{\ToneTtwoFlair} & \multicolumn{2}{c}{\ToneFlairTtwo} & \multicolumn{2}{c}{\TtwoFlairTone} & \multicolumn{2}{c}{\TtwoFlair} & \multicolumn{2}{c}{\FlairTtwo} \\ 
\cmidrule(lr){3-4} \cmidrule(lr){5-6} \cmidrule(lr){7-8} \cmidrule(lr){9-10} \cmidrule(lr){11-12}
 &  & PSNR & SSIM & PSNR & SSIM & PSNR & SSIM & PSNR & SSIM & PSNR & SSIM \\ \hline

\multirow{3}{*}{\rotatebox[origin=c]{90}{\textit{Conv}}}
& medSynth
& 25.69$\pm$2.25 & \textbf{0.911$\pm$0.031}
& 25.16$\pm$1.79 & 0.894$\pm$0.030
& 23.46$\pm$3.17 & 0.902$\pm$0.032
& 25.07$\pm$2.26 & 0.889$\pm$0.034
& 24.77$\pm$1.70 & 0.890$\pm$0.030 \\ \cline{2-12}
& pGAN
& 25.50$\pm$2.48 & 0.905$\pm$0.031
& 25.47$\pm$1.82 & 0.898$\pm$0.029
& 23.01$\pm$4.15 & 0.895$\pm$0.036
& 24.52$\pm$2.33 & 0.891$\pm$0.033
& 25.11$\pm$1.86 & 0.898$\pm$0.030 \\ \cline{2-12}
& WDM
& 25.70$\pm$2.36 & 0.901$\pm$0.032
& 25.70$\pm$1.85 & 0.903$\pm$0.028
& 23.37$\pm$3.53 & 0.900$\pm$0.033
& 25.03$\pm$2.40 & 0.893$\pm$0.034
& 25.06$\pm$1.92 & 0.896$\pm$0.029 \\ \hline

\multirow{2}{*}{\rotatebox[origin=c]{90}{\textit{Diff}}}
& DDPM
& 24.67$\pm$2.31 & 0.816$\pm$0.029
& 25.50$\pm$1.75 & 0.902$\pm$0.022
& 22.49$\pm$3.44 & 0.800$\pm$0.031
& 24.07$\pm$2.28 & 0.896$\pm$0.032
& 25.15$\pm$1.61 & 0.893$\pm$0.024 \\ \cline{2-12}
& SynDiff
& 25.70$\pm$1.68 & 0.903$\pm$0.022
& 25.14$\pm$1.53 & 0.903$\pm$0.024
& 23.01$\pm$3.77 & 0.891$\pm$0.037
& 23.59$\pm$2.22 & 0.901$\pm$0.032
& 24.82$\pm$2.26 & 0.892$\pm$0.024 \\ \hline

\multirow{4}{*}{\rotatebox[origin=c]{90}{\textit{Trans}}}
& PTNet
& 25.03$\pm$2.31 & 0.888$\pm$0.038
& 24.52$\pm$1.70 & 0.886$\pm$0.032
& 20.94$\pm$3.66 & 0.871$\pm$0.044
& 24.45$\pm$2.08 & 0.882$\pm$0.037
& 23.89$\pm$1.63 & 0.876$\pm$0.033 \\ \cline{2-12}
& ResViT
& 25.79$\pm$2.16 & 0.900$\pm$0.031
& 25.29$\pm$1.74 & 0.904$\pm$0.027
& 23.06$\pm$3.48 & 0.899$\pm$0.032
& 24.95$\pm$2.11 & 0.883$\pm$0.034
& 24.89$\pm$1.72 & 0.888$\pm$0.028 \\ \cline{2-12}
& TransUNet
& 25.54$\pm$2.24 & 0.899$\pm$0.032
& 25.59$\pm$1.86 & 0.909$\pm$0.024
& 23.26$\pm$3.52 & \textbf{0.907$\pm$0.033}
& 25.28$\pm$2.26 & 0.896$\pm$0.032
& 25.04$\pm$1.84 & 0.899$\pm$0.029 \\ \cline{2-12}
& Swin-Unet
& 25.03$\pm$2.14 & 0.887$\pm$0.035
& 24.70$\pm$1.90 & 0.886$\pm$0.032
& 22.95$\pm$3.31 & 0.887$\pm$0.037
& 24.51$\pm$2.54 & 0.879$\pm$0.034
& 24.33$\pm$1.74 & 0.881$\pm$0.033 \\ \hline

\multirow{2}{*}{\rotatebox[origin=c]{90}{\textit{SSM}}}
& U-Mamba
& 25.37$\pm$2.39 & 0.891$\pm$0.037
& 25.58$\pm$1.70 & 0.909$\pm$0.026
& 23.36$\pm$3.59 & 0.895$\pm$0.035
& 24.86$\pm$1.55 & 0.882$\pm$0.022
& 25.03$\pm$1.33 & 0.894$\pm$0.024 \\ \cline{2-12}
& Mamba-Unet
& 25.17$\pm$1.93 & 0.892$\pm$0.033
& 25.16$\pm$1.83 & 0.893$\pm$0.030
& 23.12$\pm$2.28 & 0.889$\pm$0.036
& 24.83$\pm$2.10 & 0.883$\pm$0.037
& 24.51$\pm$1.93 & 0.889$\pm$0.027 \\ \hline

& I2I-Mamba
& \textbf{26.51$\pm$2.37} & \textbf{0.911$\pm$0.033}
& \textbf{26.38$\pm$2.18} & \textbf{0.913$\pm$0.029}
& \textbf{23.83$\pm$3.79} & 0.904$\pm$0.034
& \textbf{25.79$\pm$2.44} & \textbf{0.903$\pm$0.034}
& \textbf{25.99$\pm$2.13} & \textbf{0.908$\pm$0.030} \\ \hline
\end{tabular}
}
\label{tab:brats}
\end{table*}

\begin{figure*}[t]
\centering
\includegraphics[width=0.725\textwidth]{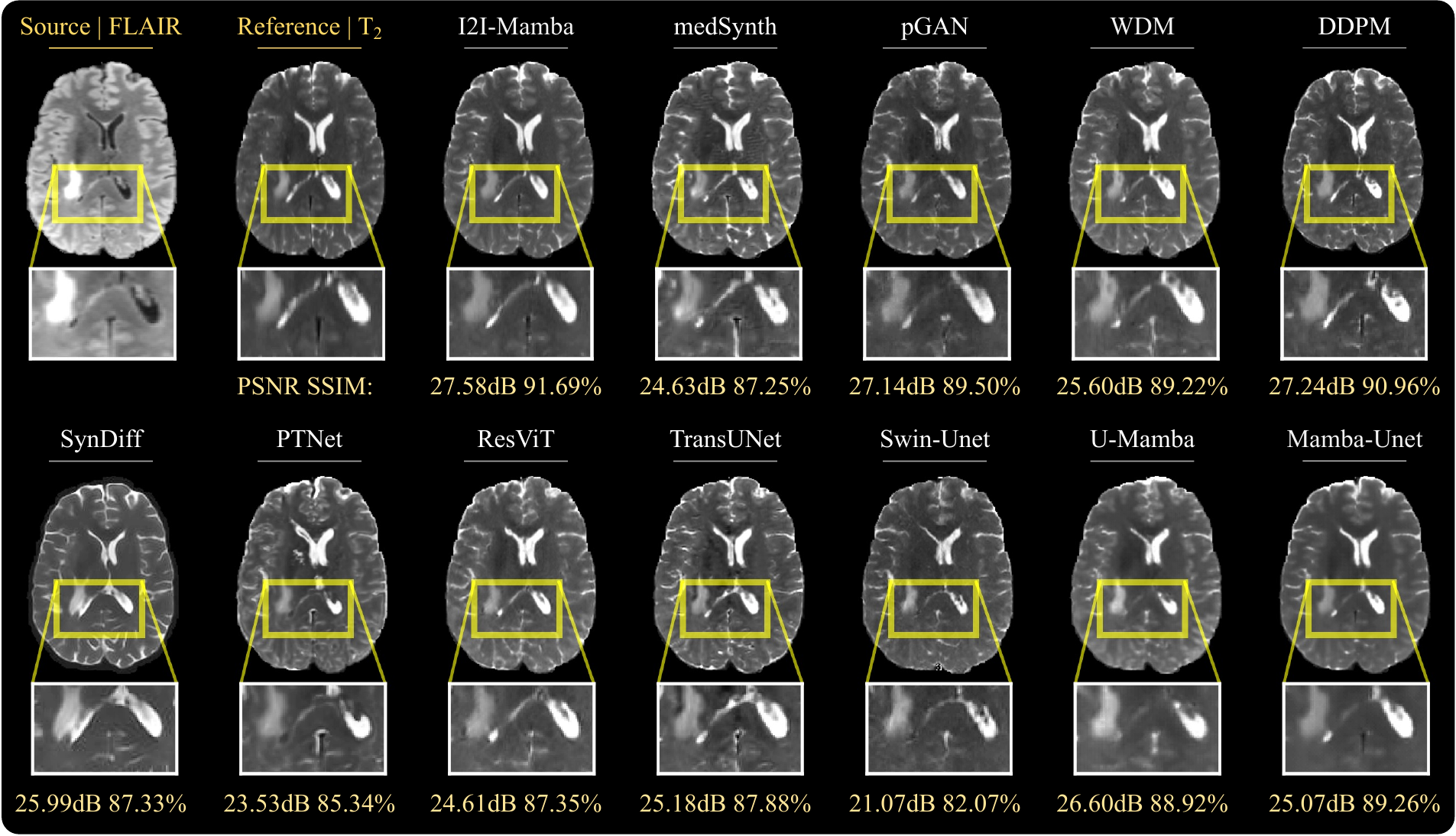}
\captionsetup{justification=justified,singlelinecheck=false}
\caption{Representative results for \FlairTtwo~in BraTS. Synthetic target images from competing methods are displayed along with source images and reference target images. Zoom-in windows and performance metrics are also included to highlight differences.}
\label{fig:BraTS}
\end{figure*}

\section{Results}
\subsection{Multi-Contrast MRI Synthesis}
We first conducted experiments for target-modality imputation in multi-contrast MRI. I2I-Mamba was comparatively demonstrated against convolutional (medSynth, pGAN, WDM), diffusion (DDPM, SynDiff), transformer (PTNet, ResViT, TransUNet, Swin-Unet), and SSM (U-Mamba, Mamba-Unet) models. Table \ref{tab:ixi} lists performance metrics for synthesis tasks in IXI that comprises data from healthy subjects. I2I-Mamba achieves the highest performance metrics consistently across tasks (p$<$0.05). On average, I2I-Mamba offers performance improvements of 2.1dB PSNR, 1.7$\%$ SSIM over convolutional baselines including the wavelet-domain WDM method; 3.0dB PSNR, 6.0\% SSIM over diffusion baselines; 2.1dB PSNR, 1.5\% SSIM over transformer baselines; and 3.2dB PSNR, 1.8\% SSIM over SSM baselines. Meanwhile, Table \ref{tab:brats} lists performance metrics for synthesis tasks in BraTS that comprises data from glioma subjects. I2I-Mamba again achieves the highest performance metrics in all tasks (p$<$0.05), except for \ToneTtwoFlair~where medSynth yields similar SSIM and \TtwoFlairTone~where TransUNet yields modestly higher SSIM. On average, I2I-Mamba offers performance improvements of 0.9dB PSNR, 1.0$\%$ SSIM over convolutional baselines; 1.3dB PSNR, 2.8\% SSIM over diffusion baselines; 1.3dB PSNR, 1.8\% SSIM over transformer baselines; and 1.0dB PSNR, 1.7\% SSIM over SSM baselines.

Synthetic target images from representative tasks are displayed in Fig. \ref{fig:IXI} for IXI, and in Fig. \ref{fig:BraTS} for BraTS. Among competing methods, convolutional baselines suffer from residual noise (e.g., pGAN) or structural inaccuracies (e.g., medSynth, WDM); diffusion baselines suffer from inaccurate depiction of low-to-moderate contrast tissue structures (e.g., DDPM) or a degree of spatial blur (e.g., SynDiff); and transformer and SSM baselines suffer either from a degree of spatial blur and contrast loss (e.g., PTNet, U-Mamba, Mamba-UNet) or pixel intensity artifacts (ResViT, TransUNet, Swin-Unet) that lead to inaccuracies in depiction of detailed anatomical structures. Particularly evident in BraTS images containing pathology, baselines generally suffer from hallucinatory features manifesting as hypo-intense or hyper-intense signals deviating from ground truth. In comparison, I2I-Mamba synthesizes target images with more accurate depiction of detailed structure and contrast in tissues, along with lower artifacts. 

Performance benefits of I2I-Mamba over SSM baselines can be attributed to the fundamental architectural differences between methods. Note that U-Mamba and Mamba-Unet follow a UNet-style architecture that substantially lowers spatial resolution of encoded feature maps, and they use conventional image-domain SSM operators based on raster-scan trajectories. In contrast, I2I-Mamba adopts a residual architecture where the bottleneck maintains relatively higher-resolution semantic representations, and it uses ddMamba blocks operating in image and Fourier domains, equipped with SSM operators based on spiral-scan trajectories. Our results indicate that these technical elements enable I2I-Mamba to sensitively capture of a comprehensive set of contextual features to synthesize high-quality target images in multi-contrast MRI protocols.

\begin{table*}[t]
\centering
\caption{Performance for the T$_2$-MRI $\rightarrow$ CT and T$_1$-MRI $\rightarrow$ CT synthesis tasks in the MRI-CT dataset.}
\resizebox{1.8\columnwidth}{!}{%
\begin{tabular}{cccccccccccccc}\hline
\multicolumn{2}{c}{} & \multicolumn{3}{c}{\textit{Conv}} & \multicolumn{2}{c}{\textit{Diff}} & \multicolumn{4}{c}{\textit{Trans}} & \multicolumn{2}{c}{\textit{SSM}} & \textit{Proposed} \\
\cmidrule(lr){3-5} \cmidrule(lr){6-7} \cmidrule(lr){8-11} \cmidrule(lr){12-13} \cmidrule(lr){14-14}
& & medSynth & pGAN  & WDM & DDPM & SynDiff & PTNet & ResViT & TransUNet & Swin-Unet & U-Mamba & Mamba-Unet & I2I-Mamba \\ \hline
\multirow{4}{*}{\rotatebox[origin=c]{90}{T$_2$ $\rightarrow$ CT}}\hspace{2mm} 
& \multirow{2}{*}{\rotatebox[origin=c]{90}{PSNR}} &  26.87 & 24.64 & 26.83 & 26.83 & 26.78 & 27.35 & 27.87 & 28.06 & 24.14 & 26.32 & 24.66 & \textbf{28.47} \\
& & $\pm$1.79 & $\pm$1.59 & $\pm$1.80 & $\pm$1.80 & $\pm$1.95 & $\pm$2.23 & $\pm$2.06 & $\pm$2.20 & $\pm$1.87 & $\pm$1.77 & $\pm$2.69 & \textbf{$\pm$2.15}\\ \cline{2-14} 
& \multirow{2}{*}{\rotatebox[origin=c]{90}{SSIM}} &  0.886 & 0.867 & 0.906 & 0.906 & 0.901 & 0.911 & 0.913 & 0.915 & 0.874 & 0.902 & 0.880 & \textbf{0.916} \\
& & $\pm$0.026 & $\pm$0.030 & $\pm$0.021 & $\pm$0.021 & $\pm$0.023 & $\pm$0.026 & $\pm$0.027 & $\pm$0.021 & $\pm$0.028 & $\pm$0.022 & $\pm$0.031 & \textbf{$\pm$0.024}\\ \hline
\multirow{4}{*}{\rotatebox[origin=c]{90}{T$_1$ $\rightarrow$ CT}}\hspace{2mm} 
& \multirow{2}{*}{\rotatebox[origin=c]{90}{PSNR}} &  27.26 & 26.92 & 26.94 & 26.70 & 27.19 & 25.61 & 26.75 & 26.55 & 25.14 & 26.38 & 23.72 & \textbf{28.01}\\
& & $\pm$2.54 & $\pm$2.51 & $\pm$1.93 & $\pm$2.59 & $\pm$1.94 & $\pm$1.98 & $\pm$1.63 & $\pm$2.01 & $\pm$2.79 & $\pm$1.85 & $\pm$2.01 & \textbf{$\pm$1.09} \\ \cline{2-14} 
& \multirow{2}{*}{\rotatebox[origin=c]{90}{SSIM}} &  0.891 & 0.904 & 0.899 & 0.905 & 0.896 & 0.863 & 0.888 & 0.900 & 0.884 & 0.894 & 0.864 & \textbf{0.909} \\
& & $\pm$0.030 & $\pm$0.029 & $\pm$0.027 & $\pm$0.026 & $\pm$0.024 & $\pm$0.025 & $\pm$0.028 & $\pm$0.021 & $\pm$0.026 & $\pm$0.030 & $\pm$0.032 & \textbf{$\pm$0.020}\\ \hline
\end{tabular}}
\label{tab:mr_ct}
\end{table*}

\begin{figure*}[t!]
\centerline{\includegraphics[width=0.75\textwidth]{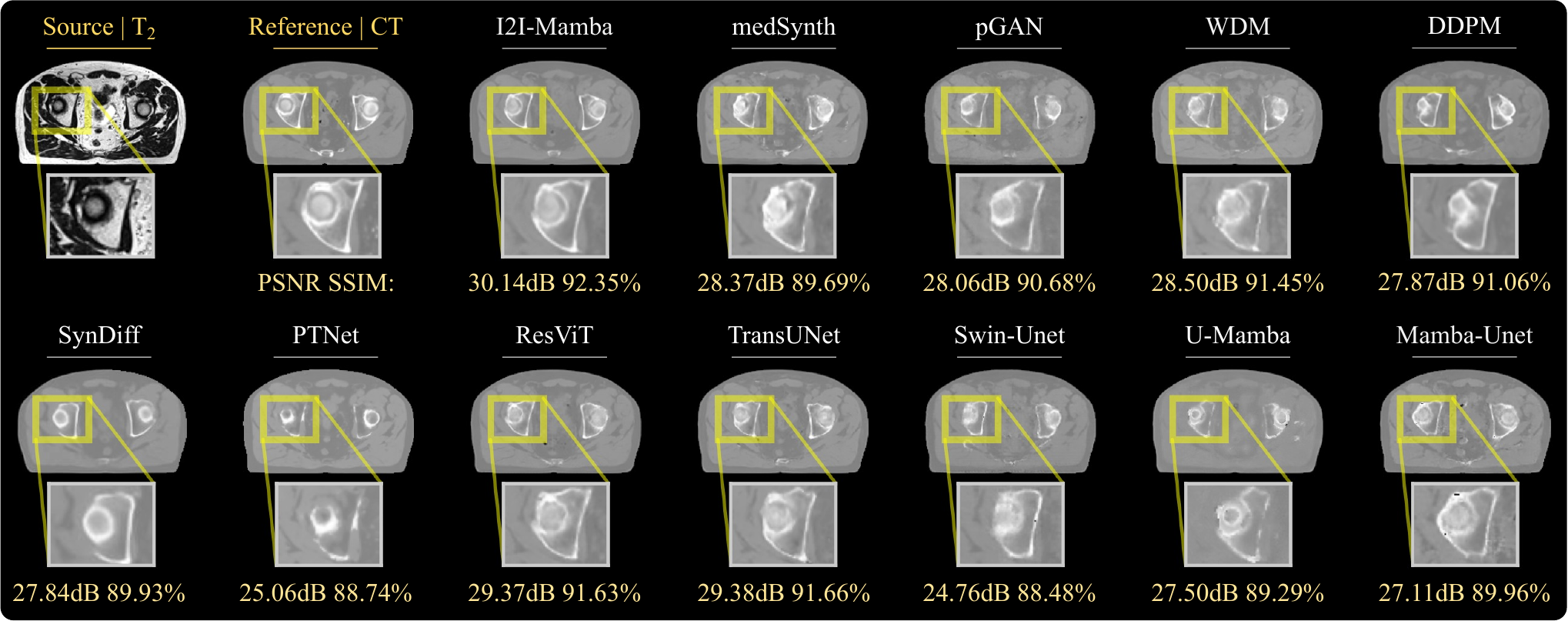}}
\caption{Representative results for \Ttwo $\rightarrow$ CT in the MRI-CT dataset. Synthetic target images from competing methods are displayed along with source images and reference target images. Zoom-in windows and performance metrics are included to highlight differences.}
\label{fig:mr_ct}
\end{figure*}

\subsection{MRI-CT Synthesis}
Next, we conducted experiments for target-modality imputation on the multi-modal MRI-CT dataset as listed in Table \ref{tab:mr_ct}. Among competing methods, I2I-Mamba achieves the highest performance metrics across tasks (p$<$0.05). On average, I2I-Mamba attains improvements of 1.7dB PSNR, 2.0$\%$ SSIM over convolutional baselines; 1.4dB PSNR, 1.1\% SSIM over diffusion baselines; 1.8dB PSNR, 1.9\% SSIM over transformer baselines; and 3.0dB PSNR, 2.8\% SSIM over SSM baselines. Consistent with the findings in multi-contrast MRI synthesis, we observe that I2I-Mamba outperforms both the wavelet-domain convolutional baseline (WDM) and the SSM-based baselines (U-Mamba, Mamba-UNet), which rely on conventional image-domain SSM operators with raster-scan trajectories. These results collectively suggest that the dual-domain SSM operators in I2I-Mamba—operating jointly in the image and Fourier domains—enhance contextual sensitivity beyond what is achievable with wavelet-domain processing or standard image-domain SSMs. This elevated sensitivity to task-relevant contextual features contributes to the improved reliability of I2I-Mamba in multi-modal synthesis.

\begin{table}[t]
\centering
\caption{Average inference times (Inf. in msec), memory load (Mem. in gigabytes) per cross-section and total number of parameters (Param. in millions) for competing methods.}
\captionsetup{justification=justified,singlelinecheck=false}
\resizebox{0.75\columnwidth}{!}{%
\begin{tabular}{ccccc}
\hline
&  & Inf. (ms) & Mem. (GB) & Param. (M) \\
\hline
\multirow{3}{*}{\rotatebox[origin=c]{90}{\textit{Conv}}} 
& medSynth & 12 & 2.9 & 76.3 \\ \cline{2-5}
& pGAN & 9 & 2.3 & 54.4 \\ \cline{2-5}
& WDM & 11 & 2.4 & 96.7 \\ \hline
\multirow{2}{*}{\rotatebox[origin=c]{90}{\textit{Diff}}} 
& DDPM & 2830 & 13.9 & 164.3 \\ \cline{2-5}
& SynDiff & 65 & 14.9 & 156.1 \\ \hline
\multirow{4}{*}{\rotatebox[origin=c]{90}{\textit{Trans}}} 
& PTNet & 34 & 10.8 & 280.4 \\ \cline{2-5}
& ResViT & 15 & 4.6 & 218.0 \\ \cline{2-5}
& TransUNet & 12 & 4.0 & 105.3 \\ \cline{2-5}
& Swin-Unet & 14 & 3.3 & 141.3 \\ \hline
\multirow{2}{*}{\rotatebox[origin=c]{90}{\textit{SSM}}} 
& U-Mamba & 10 & 1.6 & 127.8 \\ \cline{2-5}
& Mamba-Unet & 12 & 2.8 & 111.6 \\ \hline
& I2I-Mamba & 11 & 2.5 & 105.1 \\ \hline
\end{tabular}
}
\label{tab:complexity}
\end{table}

Representative synthetic images are displayed in Fig. \ref{fig:mr_ct}. Among competing methods, convolutional baselines suffer from residual noise (e.g., pGAN, WDM) or structural degradations (e.g., medSynth); and diffusion baselines suffer from structural inaccuracies in depiction of low-to-moderate contrast tissue signals (e.g., DDPM) and a degree of spatial blur (e.g., SynDiff). Meanwhile, transformer and SSM baselines suffers from spatial smoothing (e.g., ResViT, TransUNet, Swin-Unet, U-Mamba), hypo- or hyper-intense contrast compared to ground truth in regions near bone tissue (e.g., PTNet, U-Mamba), or dark-pixel artifacts (e.g., Swin-Unet, Mamba-Unet). In comparison, I2I-Mamba synthesizes target images with lower artifacts and more accurate delineation of tissue structure and contrast, particularly across diagnostically relevant bone regions. Taken together, these results indicate that I2I-Mamba maintains higher sensitivity to contextual features in multi-modal medical images than baselines, thereby increasing fidelity in imputation of missing modalities.

\subsection{Computational Complexity}
Inference times and memory load per cross-section for all competing methods are summarized in Table \ref{tab:complexity} along with model parameter counts. As expected, both diffusion and transformer models exhibit substantial computational overhead, reflected in longer inference times (due to iterative image generation in diffusion, and high model complexity in transformer models), along with higher memory usage and increased parameter counts. This trend holds for hybrid CNN-transformer architectures (e.g., ResViT, TransUNet) as well as more efficient transformer variants employing approximate attention mechanisms (e.g., PTNet, Swin-Unet). In contrast, convolutional baselines offer high computational efficiency, with pure image-domain architectures (e.g., pGAN, medSynth) demonstrating the lowest inference times and memory demands. SSM-based methods, including I2I-Mamba, exhibit notably higher efficiency than transformer methods and show generally competitive efficiency to convolutional methods. Note that I2I-Mamba achieves inference time and memory usage comparable to the most efficient convolutional baselines, while its parameter count, though higher than pGAN and medSynth, remains on par with the wavelet-domain baseline (WDM). Taken together with its superior synthetic image quality, these results highlight I2I-Mamba as an efficient and effective solution for multi-modal medical image synthesis.

\subsection{Ablation Studies}
We performed a systematic set of ablation studies on I2I-Mamba to assess the contribution of its key design elements and configurations to synthesis performance. 

\textbf{Architectural components:} First, we examined the importance of SSM operators in image and Fourier domains to capture spatial and spectral context, channel-mixing layers to aggregate context across the channel dimension, and rCNN blocks to enhance spatial precision. I2I-Mamba was compared against several variants for this purpose: `w/o ddMamba' that ablated ddMamba blocks entirely, `w/o ddMamba$^F$' that ablated the Fourier-domain branch from the ddMamba blocks, `w/o ddMamba$^I$' that ablated the image-domain branch from the ddMamba blocks,`w/o chan mix' that ablated channel-mixing layers from ddMamba blocks, and `w/o rCNN' that ablated rCNN blocks. Table \ref{tab:ablation_1} lists performance metrics for all variants in representative synthesis tasks. We find that I2I-Mamba consistently outperforms all variants (p$<$0.05), with improvements up to 1.0dB PSNR and 1.4\% SSIM. These results indicate that each examined design element in I2I-Mamba contributes significantly to model performance.

\begin{table}[t]
\centering
\caption{Performance of I2I-Mamba variants built by: ablating ddMamba blocks (w/o ddMamba), removing the Fourier-domain branch from the ddMamba blocks (w/o ddMamba$^F$), removing the image-domain branch from the ddMamba blocks (w/o ddMamba$^I$), ablating channel-mixing layers (w/o chan mix), and ablating rCNN blocks (w/o rCNN). Results listed for representative synthesis tasks of \ToneTtwoPD\,\,in IXI, \FlairTtwo\,\,in BraTS, and \Tone\,$\rightarrow$\,CT in the MRI-CT dataset.}
\captionsetup{justification = justified,singlelinecheck = false}
\resizebox{0.9\columnwidth}{!}{%
\begin{tabular}{cccccccc}
\hline
\multirow{2}{*}{} & \multicolumn{2}{|c|}{\ToneTtwoPD} & \multicolumn{2}{|c|}{\FlairTtwo} & \multicolumn{2}{|c|}{\Tone $\rightarrow$ CT} \\ \cline{2-7} 
                  & PSNR      & SSIM     & PSNR      & SSIM      & PSNR      & SSIM     \\ \hline
\multirow{2}{*}{I2I-Mamba} & \textbf{33.46} & \textbf{0.969} & \textbf{25.99} & \textbf{0.908} & \textbf{28.01} & \textbf{0.909} \\
            & \textbf{$\pm$2.52} & \textbf{$\pm$0.011} & \textbf{$\pm$2.13} & \textbf{$\pm$0.030} & \textbf{$\pm$1.09} & \textbf{$\pm$0.020} \\ \hline
\multirow{2}{*}{w/o ddMamba} & 32.65 & 0.966 & 24.86 & 0.899 & 27.18 & 0.902 \\
            & $\pm$2.56 & $\pm$0.010 & $\pm$2.05 & $\pm$0.029 & $\pm$1.98 & $\pm$0.021  \\ \hline
\multirow{2}{*}{w/o ddMamba$^{F}$} & 32.91 & 0.967 & 25.33 & 0.902 & 27.43 & 0.904 \\
            & $\pm$2.40 & 0.012 & $\pm$1.56 & $\pm$0.032 & $\pm$1.92 & $\pm$0.035 \\ \hline
\multirow{2}{*}{w/o ddMamba$^{I}$} & 33.09 & 0.968 & 25.80 & 0.907 & 27.59 & 0.906 \\
            & $\pm$2.58 & 0.012 & $\pm$2.13 & $\pm$0.030 & $\pm$2.06 & $\pm$0.024 \\ \hline
\multirow{2}{*}{w/o chan mix} & 33.14 & 0.968 & 25.19 & 0.898 & 27.53 & 0.897 \\
            & $\pm$2.63 & $\pm$0.012 & $\pm$1.91 & $\pm$0.028 & $\pm$2.41 & $\pm$0.028 \\ \hline
\multirow{2}{*}{w/o rCNN} & 32.69 & 0.966 & 25.07 & 0.894 & 26.60 & 0.885 \\
            & $\pm$2.53 & $\pm$0.011 & $\pm$1.83 & $\pm$0.024 & $\pm$2.05 & $\pm$0.030 \\ \hline
\end{tabular}
}
\label{tab:ablation_1}
\end{table}

\renewcommand{\tabcolsep}{4pt}
\renewcommand{\arraystretch}{1.275}
\begin{table}[t]
\centering
\caption{Performance of I2I-Mamba variants built by using: 1D bidirectional sweep scans (w sweep$^\text{a}$), 2D bidirectional sweep scans (w sweep$^\text{b}$), 1D bidirectional zigzag scans (w zigzag$^\text{a}$), 2D bidirectional zigzag scans (w zigzag$^\text{b}$) and ensemble sweep-zigzag scans (w sweep$^\text{a}$-zigzag$^\text{a}$).}
\captionsetup{justification = justified,singlelinecheck = false}
\resizebox{0.85\columnwidth}{!}{%
\begin{tabular}{ccccccc}
\hline
\multirow{2}{*}{} & \multicolumn{2}{|c|}{\ToneTtwoPD} & \multicolumn{2}{|c|}{\FlairTtwo} & \multicolumn{2}{|c|}{\Tone $\rightarrow$ CT} \\ \cline{2-7} 
                  & PSNR      & SSIM     & PSNR      & SSIM      & PSNR      & SSIM     \\ \hline
\multirow{2}{*}{I2I-Mamba} & \textbf{33.46} & \textbf{0.969} & \textbf{25.99} & \textbf{0.908} & \textbf{28.01} & \textbf{0.909} \\
            & \textbf{$\pm$2.52} & \textbf{$\pm$0.011} & \textbf{$\pm$2.13} & \textbf{$\pm$0.030} & \textbf{$\pm$1.09} & \textbf{$\pm$0.020} \\ \hline
\multirow{2}{*}{w sweep$^\text{a}$} & 32.72 & 0.964 & 24.97 & 0.891 & 27.31 & 0.890\\
            & $\pm$2.48 & 0.013 & $\pm$1.86 & $\pm$0.031 & $\pm$2.11 & $\pm$0.027 \\ \hline
\multirow{2}{*}{w sweep$^\text{b}$} & 32.82 & 0.963 & 25.01 & 0.903 & 27.26 & 0.895 \\
            & $\pm$2.49 & $\pm$0.012 & $\pm$1.98 & $\pm$0.031 & $\pm$2.58 & $\pm$0.034 \\ \hline
\multirow{2}{*}{w zigzag$^\text{a}$} & 32.34 & 0.958 & 24.41 & 0.875 & 26.02 & 0.865 \\
            & $\pm$2.77 & $\pm$0.013 & $\pm$1.95 & $\pm$0.029 & $\pm$2.17 & $\pm$0.033 \\ \hline
\multirow{2}{*}{w zigzag$^\text{b}$} & 32.31 & 0.957 & 24.37 & 0.898 & 26.11 & 0.869 \\
            & $\pm$2.65 & $\pm$0.012 & $\pm$1.88 & $\pm$0.033 & $\pm$2.25 & $\pm$0.035 \\ \hline
\multirow{2}{*}{\begin{tabular}{c}w sweep$^\text{a}$-\\zigzag$^\text{a}$\end{tabular}} & 32.04 & 0.955 & 24.01 & 0.883 & 26.75 & 0.878 \\
            & $\pm$2.28 & $\pm$0.011 & $\pm$1.91 & $\pm$0.030 & $\pm$1.85 & $\pm$0.029 \\ \hline
\end{tabular}
}
\label{tab:ablation_2}
\end{table}

\begin{figure*}[t]
\centerline{\includegraphics[width=0.95\textwidth]{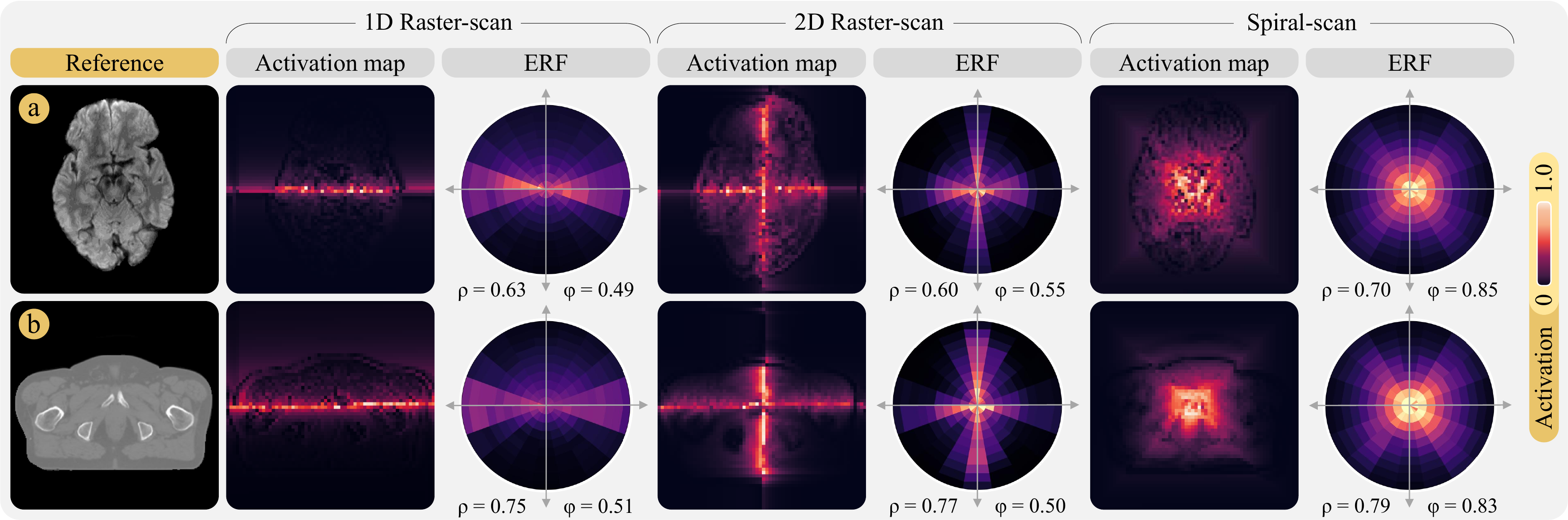}}
\caption{Per-sample activation maps and population-average effective-receptive fields (ERF) are shown for \textbf{(a)} \FlairTtwo\,\,in BraTS, \textbf{(b)} \Tone\,$\rightarrow$\,CT in MRI-CT. Depicting contextual interactions of a central pixel with remaining pixels across the image, activation maps and ERFs were extracted from the SSM layer at bottleneck stage $j$=1 (64$\times$64 resolution). Results for `w sweep$^\text{a}$' (1D bidirectional raster-scans), `w sweep$^\text{b}$' (2D bidirectional raster-scans) and I2I-Mamba (spiral-scan), along with reference target images. For ERFs, $\rho$ denotes the radial coverage, $\varphi$ denotes the angular isotropy, both metrics normalized to a maximum of 1.}
 \label{fig:attention}
\end{figure*}

\textbf{Spiral-scan SSM operator:} Next, we examined the importance of implementing SSM operators based on the proposed spiral-scan trajectory. I2I-Mamba was compared against several variants based on raster-scan trajectories \cite{liu2024vmamba}: `w sweep$^\text{a}$' that used 1D bidirectional sweep scans (2 separate scan trajectories), `w sweep$^\text{b}$' that used 2D bidirectional sweep scans (4 trajectories), `w zigzag$^\text{a}$' that used 1D bidirectional zigzag scans, `w zigzag$^\text{b}$' that used 2D bidirectional zigzag scans, and `w sweep$^\text{a}$-zigzag$^\text{a}$' that used an ensemble of sweep and zigzag scans. As listed in Table \ref{tab:ablation_2}, we find that I2I-Mamba outperforms all variants consistently across tasks (p$<$0.05), with improvements up to 1.6dB PSNR and 2.3\% SSIM.

\textbf{Assessment of ERFs:} To visually assess the influence of the spiral-scan trajectory on the SSM operator footprint, we extracted per-sample activation maps and population-average effective receptive fields (ERFs) for SSM layers \cite{liu2024vmamba}. Fig. \ref{fig:attention} displays activation maps and ERFs in representative cross-sections for I2I-Mamba (i.e., spiral scan), `w sweep$^\text{a}$' (i.e., 1D bidirectional raster scans), and `w sweep$^\text{b}$' (i.e., 2D bidirectional raster scans). Note that since latent representations in SSMs are processed in a sequence-dependent manner, regions of high activation may reflect abstract contextual dependencies rather than detailed anatomical correspondences in the input image. We observe that I2I-Mamba significantly improves angular isotropy of the SSM operator while maintaining coverage over a broad neighborhood of pixels. In contrast, `w sweep$^\text{a}$', `w sweep$^\text{b}$' suffer from notable bias across the primary scan directions (i.e., horizontal and/or vertical), diminishing focus over regions distanced obliquely to the scan directions. 

For quantitative assessment, ERFs were transformed into a polar representation, partitioned into 10 radial bins and 20 angular wedges. Radial coverage ($\rho$) was measured as the percentage of radial distances with normalized activation values above 0.1 \cite{Fisher_1993}. Angular isotropy ($\varphi$) was measured as $1 - CV$ across orientations where $CV$ denotes coefficient of variation \cite{Fisher_1993}. Higher $\rho$ reflects greater spatial spread in the ERF, hence offering enhanced sensitivity to long-range context. Meanwhile, higher $\varphi$ reflects increased directional uniformity in the ERF, mitigating bias against non-axial orientations. On average across the test sets, I2I-Mamba's ERFs show moderately higher radial coverage $\rho$$\,=\,$0.75 compared to `w sweep$^\text{a}$' with $\rho$$\,=\,$0.69 and `w sweep$^\text{b}$' with $\rho$$\,=\,$0.69 (p$<$0.05). Importantly, I2I-Mamba's ERFs show substantially enhanced angular isotropy $\varphi$$\,=\,$0.84 than both `w sweep$^\text{a}$' with $\varphi$$\,=\,$0.50 and `w sweep$^\text{b}$' with $\varphi$$\,=\,$0.53 (p$<$0.05). Note that I2I-Mamba sustains these benefits using only two spiral scans (one scan in each branch of ddMamba blocks), compared to the four scans in `w sweep$^\text{b}$'. These results indicate that the enhanced footprint of I2I-Mamba’s spiral-scan SSM operator improves its capacity to capture a more comprehensive set of contextual features compared to conventional SSM operators. Furthermore, the observed performance gains, along with the improved ERF attributes, suggest that I2I-Mamba effectively captures semantically meaningful content in latent representations of medical images.

\textbf{Injection of ddMamba blocks:} Lastly, we examined the importance of injecting ddMamba blocks in select stages of the bottleneck (i.e., $S=\{b1,b5,b9\}$). To do this, we compared I2I-Mamba against variants built by inserting ddMamba blocks in different configurations across the bottleneck, as well as across the encoder-decoder as listed in Table \ref{tab:ablation_3}. While all models generally yield similar SSIM values, we find that I2I-Mamba consistently yields superior PSNR against variants (p$<$0.05). Lower performance in variants with fewer ddMamba blocks can be attributed to a lowered ability to extract contextual features, whereas lower performance in variants with a greater number of ddMamba blocks is best attributed to elevated model complexity. These results suggest that I2I-Mamba attains a favorable trade-off between contextual sensitivity and model complexity by avoiding indiscriminate placement.

\renewcommand{\tabcolsep}{4pt}
\renewcommand{\arraystretch}{1.25}
\begin{table}[t]
\centering
\caption{Performance of I2I-Mamba variants obtained with different configurations of inserting ddMamba blocks ($S$). $S^*$=\{b1,b5,b9\} denotes the configuration reported in the main experiments. $e$, $b$, $d$ respectively denote encoder, bottleneck, and decoder stages.}
\captionsetup{justification = justified,singlelinecheck = false}
\resizebox{0.95\columnwidth}{!}{%
\begin{tabular}{ccccccc}
\hline
\multirow{2}{*}{} & \multicolumn{2}{|c|}{\ToneTtwoPD} & \multicolumn{2}{|c|}{\FlairTtwo} & \multicolumn{2}{|c|}{\Tone $\rightarrow$ CT} \\ \cline{2-7} 
                  & PSNR      & SSIM     & PSNR      & SSIM      & PSNR      & SSIM     \\ \hline
\multirow{2}{*}{$S$=\{b1,b2,...,b9\}} & 33.18 & \textbf{0.969} & 25.72 & 0.905 & 27.50 & 0.906 \\
            & $\pm$2.50 & \textbf{$\pm$0.011} & $\pm$2.00 & $\pm$0.029 & $\pm$1.90 & $\pm$0.024 \\ \hline
\multirow{2}{*}{$S$=\{b1,b3,b5,b7,b9\}} & 33.14 & \textbf{0.969} & 25.83 & 0.907 & 27.62 & \textbf{0.909} \\
            & $\pm$2.58 & \textbf{$\pm$0.011} & $\pm$2.07 & $\pm$0.028 & $\pm$2.28 & \textbf{$\pm$0.026} \\ \hline
\multirow{2}{*}{$S^*$=\{b1,b5,b9\}} & \textbf{33.46} & \textbf{0.969} & \textbf{25.99} & \textbf{0.908} & \textbf{28.01} & \textbf{0.909} \\
            & \textbf{$\pm$2.52} & \textbf{$\pm$0.011} & \textbf{$\pm$2.13} & \textbf{$\pm$0.030} & \textbf{$\pm$1.09} & \textbf{$\pm$0.020} \\ \hline
\multirow{2}{*}{$S$=\{e1-e3, b1,b5,b9, d1-d3\}} & 32.76 & 0.957 & 25.87 & \textbf{0.908} & 28.00 & 0.908 \\
            & $\pm$2.76 & $\pm$0.014 & $\pm$1.92 & \textbf{$\pm$0.028} & $\pm$2.01 & $\pm$0.025 \\ \hline
\multirow{2}{*}{$S$=\{b5\}} & 33.09 & \textbf{0.969} & 25.94 & \textbf{0.908} & 27.84 & 0.907 \\
            & $\pm$2.45 & \textbf{$\pm$0.013} & $\pm$2.01 & \textbf{$\pm$0.027} & $\pm$2.09 & $\pm$0.024 \\ \hline
\end{tabular}
}
\label{tab:ablation_3}
\end{table}

\section{Discussion}
\subsection{Scope and Implications}
In this work, we introduced I2I-Mamba, a novel state-space model (SSM)-based model for imputing missing modalities in multi-modal medical imaging. Conventional CNNs lack sensitivity to long-range dependencies across distant anatomical regions \cite{ptnet}, while transformer models, though better at modeling such dependencies, suffer from quadratic complexity that restricts attention to low-resolution or coarse features \cite{trans_unet, resvit}, limiting their contextual sensitivity in high-resolution image synthesis tasks. SSMs have recently emerged as an efficient alternative, yet existing variants typically operate in the image domain with raster-scan trajectories, which constrain their capture of global and frequency-specific context and introduce directional bias in their receptive fields across non-axial orientations. I2I-Mamba addresses these limitations via dual-domain spiral-scan SSM operators that jointly process features in the image and Fourier domains, enhancing contextual sensitivity without incurring high computational costs. Our results show that I2I-Mamba consistently outperforms CNN, transformer, and prior SSM baselines in both quantitative accuracy and visual quality. 

I2I-Mamba's performance benefits might be of value in several clinical scenarios where efficient, high-fidelity synthesis of missing modalities is needed: (i) imputing redundant but diagnostically useful target modalities to reduce total scan time; (ii) inferring invasive or high-risk modalities—such as those requiring contrast agents or ionizing radiation—using safer, non-invasive scans \cite{lee2019}; and (iii) retrospectively recovering missing modalities to enhance harmonization across large-scale imaging datasets in population studies \cite{divbar2019}.

\subsection{Limitations and Future Work}
Several lines of limitations can be addressed to help further improve the proposed method's performance. A first group of developments concerns learning strategies. Here we examined one-to-one and many-to-one tasks to impute missing images in multi-contrast MRI and MRI-CT protocols. For optimal performance, a separate model was built for each individual task. In certain scenarios, missing and acquired modalities in a multi-modal protocol may vary sporadically across the imaging cohort \cite{mmgan}. To improve practicality, a unified I2I-Mamba can be built by adopting a masked training strategy so as to perform many-to-many synthesis tasks \cite{mmgan,resvit}. When needed, reliability of such multi-tasking models might be boosted via multi-site datasets that would provide access to larger and more diverse sets of training samples \cite{pFLSynth}. Here, we performed supervised learning by assuming that paired sets of source and target modality images are available in each training subject \cite{pgan}. In cases where it is difficult to curate paired training images, unsupervised learning strategies based on cycle-consistency could be adopted to permit training on unpaired images \cite{woltering2017,ge2019}. 

A second group of developments concerns the synthesis tasks implemented. Corroborating recent findings, here we observed high performance in translation among endogenous MRI contrasts and in translation from MRI to CT. Literature suggests that synthesizing exogenous MRI contrasts that involve external contrast agents, or synthesis of MRI images from CT are rather ill-posed problems \cite{lee2019,resvit}. Incorporating regularization priors regarding the distribution of the target modality might help improve synthesis fidelity in such challenging tasks \cite{yu2019ea,lee2017}. For effective yet practically scalable synthesis, we demonstrated our method on two-dimensional images, consistent with prevailing practices in recent literature. Although our approach readily supports multi-slice inputs for processing volumetric slabs, we focused on cross-sectional evaluations to strike a favorable balance between accuracy and computational efficiency \cite{pgan}. Notably, we did not observe any intensity consistency issues across slices, suggesting that explicit volumetric modeling offers no significant advantage for the tasks considered \cite{ptnet}. Investigating the potential benefits of state-space models in volumetric synthesis remains an important direction for future work.

A third group of developments concerns the employed loss functions. For systematic evaluations under an efficient framework, here we implemented primary competing methods based on a combined pixel-wise/adversarial loss \cite{resvit}. This adversarial learning approach elicited high-quality synthetic images in the reported experiments, and in preliminary studies we did not observe a notable benefit from a diffusion-based implementation of I2I-Mamba. That said, further improvements in image quality might be viable with advanced loss terms including gradient-based, difficulty-aware, cross-entropy losses \cite{nie2018,Luping1,Shen1}. A potential limitation of adversarial learning is instabilities in model training that can hamper the fidelity of synthetic images \cite{syndiff}. While here we did not observe signs of instability in model training, diffusion learning could be adopted to improve reliability when necessary \cite{SelfRDB,vmddpm}. Additional improvements to be sought for a diffusion-based implementation include bridge formulations to boost task-relevant information \cite{ssdiffrecon}. Lastly, synthesis performance might be further improved by adopting model pre-training procedures. Future studies are warranted for an in-depth evaluation of the ideal training procedures for I2I-Mamba.  

\section{Conclusion}
In this study, we proposed a novel learning-based method for imputing missing modalities in multi-modal medical imaging protocols. To improve fidelity in synthetic images, I2I-Mamba leverages a novel SSM-based architecture with dual-domain Mamba blocks operating in image and Fourier domains to capture spatial and spectral contextual features. These ddMamba blocks are further equipped with spiral-scan trajectories to improve angular isotropy in the receptive field of SSM operators. With its superior performance and competitive computational efficiency with respect to state-of-the-art baselines, I2I-Mamba holds great potential for multi-modal medical image synthesis.

\bibliographystyle{IEEEtran}
\bibliography{refs}

\end{document}